# Genuine Plane Symmetries versus Pseudosymmetries in two Crystal Patterns of Graphic Artwork

Peter Moeck

Department of Physics, Portland State University, Portland, OR, USA; pmoeck@pdx.edu

**Abstract**

The reader is informed about a method for the objective identification of the plane symmetry group of a "noisy" crystal pattern. Without giving numerical details, this information theory based method is applied to two beautiful pieces of graphic art. The plane symmetry group identifications distinguish between genuine symmetries and pseudosymmetries as a byproduct. Pieces of graphic/geometric artworks are ideal for the further refinement of the new method because they are macroscopic and their "noise content" is chiefly due to the individual artists' handiwork and employed creative procedures. As different graphic techniques/procedures were employed in the creation of the classified crystal patterns, one may glean insights on how well a particular technique or procedure supports the realization of a crystallographic symmetry group in a graphic work of art.

## Introduction

The author of this paper has developed *objective* methods for the classification of the whole range of crystallographic symmetries of noisy micrographs that were recorded in digital form from crystals and crystal surfaces [1]. Such micrographs/images are, per definition [2], finite crystal patterns that are more or less translation periodic in two dimensions (2D). In other words, they are to be understood as noisy versions of abstract crystal patterns, which are by definition perfectly symmetric. In the following, it is understood that crystal patterns that originate from real-world physical objects are always finite and noisy. The noise resides thereby at the individual image-pixel level in the form of a measurable deviation of a pixel's actual value from its idealized value that is prescribed by the underlying abstract crystal pattern.

The objectivity of the author's methods is ensured by the adaptation of a geometric form of information theory [3] to classifications of digital image data into the crystallographic symmetries [4] of the Euclidian plane. Prior to these developments, *subjectively* defined thresholds for allowed deviations of perceived symmetries in real-world crystal patterns needed to be employed for such classifications. All symmetries in all regular real-world objects are always broken because they are abstract mathematical concepts rather than physical properties of whatever it is that is to be imaged, i.e. projected into two dimensions, for a subsequent 2D symmetry classification of its noisy image. The geometric symmetries themselves are perfect/unbroken and properties of an abstract mathematically defined space.

Digital photos of more or less 2D translation periodic pieces of graphic art can also be considered to be projections of physical objects that fall under the definition of a 2D crystal pattern. They can, therefore, be classified with respect to their plane symmetry group with one of the above-mentioned methods whenever objectivity and *quantifications* are desired.

Pseudosymmetries [5] do not distract from the beauty of graphic art but add to it. Because pseudosymmetries are not rare in nature, they are some kind of a nuisance to crystallographers and structural chemists when they refer to atomic positions in crystals and molecules. The deviation of some plane symmetries of more or less 2D periodic patterns from their respective mathematical definitions are often much smaller than the breaking of some other symmetries. (This fact is commonly used to justify the use of the above-mentioned thresholds.) Using subjectively set thresholds, the distinctions between the genuine symmetries and pseudosymmetries in a crystal pattern become, however, rather arbitrary.

The purpose of this paper is twofold. The first goal is to inform the reader about the author's objective plane symmetry group classification method. Because genuine crystallographic symmetries are now distinguishable from pseudosymmetries as a byproduct of the application of that method, the second goal of this paper is to illustrate their difference at the quantitative level. The employed definitions for pseudosymmetries and genuine symmetries are those of the online dictionary of the International Union of Crystallography. Other mathematically defined crystallographic concepts that are used in this paper can also be looked up there by clicking on links in the list of references.

The rest of the paper is organized as follows. The difference between a symmetry in a crystal pattern that needs to be labeled as either genuine symmetry or pseudosymmetry is clarified in Fig. 1, its annotations, and the associated discussion. Key features of the new method are then briefly described without recourse to mathematical equations, inequalities, and numbers. Two pieces of graphic art are subsequently classified with respect to their plane symmetry group and pseudosymmetries. A few site [6] symmetries are then assigned to points in these crystal patterns.

Because they are macroscopic, more or less translation periodic pieces of graphic artworks are ideal for the further refinement of the new method. The gray-value deviations of the individual pixel values of graphic artworks from their perfectly symmetric abstractions are there chiefly due to the individual artists' handiwork and employed creative procedures. (Contributions of the less than perfect recording of the digital images to the generalized noise [1] in a classification are minimal because no microscope needs to be involved.) The paper ends with a summary and conclusions section.

## Distinction between Genuine Symmetries and Pseudosymmetries

Figure 1 illustrates in a pictorial form the difference between genuine symmetries and pseudosymmetries at the point symmetry level. (Point symmetries that are part of crystal patterns are called site symmetries.) At the center of this figure, there is a two-fold rotation point. This is the least broken symmetry in that figure at a quantitative (rather than a qualitative) level and, therefore, per definition [1] a genuine symmetry. At first sight there seem to be two mirror lines, which intersect each other under a right angle. These lines are of an auxiliary nature and dotted in Fig. 1.

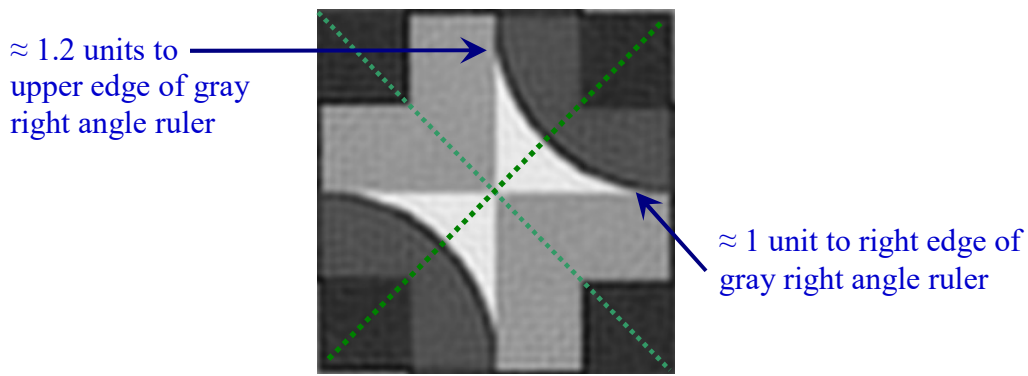

*Figure 1:* *Geometric object with genuine point symmetry group 2 and pseudosymmetries in the form of intersecting mirror lines, modified from [1].*

Closer inspection of Fig. 1 reveals that the mirror symmetry around the two auxiliary lines are on a quantitative level more severely broken than the two-fold rotation symmetry around the center of the geometric object. Because these breakings are in the information-theoretic sense [3] severely enough, the auxiliary lines denote only pseudosymmetries. (Whenever a geometric object such as the one in Fig. 1 is part of a crystal pattern, all of these symmetry breakings can be quantified and objectively ranked by one of the methods in [1].) When the two pseudosymmetries of the geometric object in Fig. 1 are combined with the genuine two-fold rotation point, pseudosymmetry point group '*2mm*' results. As will be shown below, there is no genuine *2mm* site symmetry in a charming collage by Eva Knoll (that is classified on the sixth and seventh page of this paper with respect to its plane symmetry group and pseudosymmetries).

# Key Features of the Objective Plane Symmetry Group Identification Method

The obvious advantage of working from a digital image of a graphic artwork is that the individual pixels' characteristics, i.e. their gray-level values and associated 2D image coordinates, are measured and objectively assigned quantities that can be subjected to statistical analyses. As part of such analyses, one needs to compare all gray-level pixel values of the real-world crystal pattern to their counterparts in multiple geometric models that were generated from the crystal pattern that is to be classified. Each of the geometric models of the real-world crystal pattern is a differently symmetrized version of that crystal pattern. If there is more than translation symmetry in the digital image of the crystal pattern that is to be classified, one of these abstractions will provide the best representation to said pattern whereby its geometric degrees of freedom need to be taken into account properly (employing established mathematical procedures).

Making a crystallographic symmetry classification is, thus, synonymous to solving a geometric model selection problem. Note that the translation-periodic (structural) information in a real-world crystal pattern is explicitly modeled by different symmetry abstractions from the digital image of that pattern in a crystallographic symmetry classification process. The distinction about which part of the crystal pattern would probably be perfectly symmetric (redundant) in the hypothetical case of the complete absence of all deviations and which parts contain probably only information that one is not interested in is a byproduct of the classification. That latter type of information is simply noise, or in another word "non-information," which needs to be extrapolated away in order to arrive at an abstract structural-mathematical truth.

The gray levels of the individual pixels in the image of the crystal pattern that is to be classified are considered to be the sum of a structural part, i.e. the structural/symmetric information, and a non-structural part, i.e. the noise that obscures this information. That noise is considered to be approximately Gaussian distributed in Kenichi Kanatani's geometric Akaike Information Criterion (G-AIC) [3], as that is the common approach in science and engineering in general when actual noise distributions are unknown. The main causes for these deviations are in this paper assumed to be due to the individual artists' handiwork and employed creative procedures. (This assumption is very reasonable as deviations that originate from imperfections in the imaging and digitization of graphic artworks as well as inaccuracies in calculations by the used computer programs, e.g. effects of using approximating series instead of trigonometric functions and accumulated rounding errors, are typically much smaller than the deviations that are present in the original artwork.) One may accordingly glean insights on how well the results of the utilization of a particular technique/procedure by a particular artist adhere to the realization of a crystallographic symmetry group in a particular graphic/geometric piece of art.

The selection of the most representative model for numerical data from a set of alternative models is an application of information theory. The models are here geometric in nature and many of them are necessarily non-disjoint from each other. The latter fact can be appreciated from a glance into [7] where all the maximal subgroups and minimal supergroups are listed for all of the plane symmetry groups. The non-disjointedness of the geometric models complicates the model selection process as any lower symmetric model will always fit any image data by any distance measure at the individual pixel level better than a non-disjoint higher symmetric model for the same data when only the quantified deviations of the individual pixel values from their symmetry abstractions are taken into account.

Kanatani's G-AIC overcomes this problem by balancing squared residuals between the image data and its geometric models with penalty terms that depend on the number of geometric degrees of freedom of these models. The more general, i.e. less symmetric, model possesses more geometric degrees of freedom than the more sophisticated, i.e. more symmetric, model. The less symmetric model obtains, therefore, a greater additive penalty term to its lower squared residual as part of its G-AIC value. The model with the lower G-AIC value is taken to be the better (information-theoretic) representation of the image/crystal pattern as far as its structural-periodic content is concerned.

Crystallographic symmetry classifications are best performed in Fourier/reciprocal [8] space. This is because calculating the discrete Fourier transform of the digital image of the crystal pattern that is to be classified leads to the translation averaging of that image as a byproduct. Selecting only the (periodic) structure-bearing complex Fourier coefficients of that transform as translation-averaged reciprocal-space representation of the graphic artwork leads to a significant filtering out of noise. It is also straightforward and computationally highly effective to create the geometric models of the crystal pattern that is to be classified in Fourier space. One simply has to enforce the symmetry relations and restrictions of the plane symmetry groups on the amplitude and phase angle parts of those complex Fourier coefficients that are laid out on the reciprocal lattice in the amplitude map of the transform of the crystal pattern that is to be classified. The above-mentioned squared residuals of Kanatani's G-AIC are then obtained as squares of the complex difference between the (periodic) structure-bearing Fourier coefficients of the crystal pattern that is to be classified and those of its various geometric models.

Ratios of squared residuals of non-disjoint models play an important part in Kanatani's framework. Such ratios are given in Fig. 2 for equal numbers of structure-bearing complex Fourier coefficients in both the transform of the image of the crystal pattern that is to be classified and its various geometric models. This figure shows the applicable hierarchy of the plane symmetry groups, whereby the number of non-translational symmetry operations [9] increases from the bottom to the top as organizing principle. Arrows lead from maximal subgroups to their minimal supergroups. The latter feature all of the non-translational symmetry operations of the subgroups plus one additional symmetry operation and the consequences of this addition. (Note in passing that the Hermann-Mauguin notation [10] of the plane symmetry groups in Fig. 2 reveals the maximal subgroup to minimal supergroup relationships pretty well.) One can metaphorically "climb up" from a maximal subgroup to its minimal supergroup as long as the ratio of the squared residual values of a non-disjoint model pair fulfills the inequality that is applicable for a transition from a lower level of the hierarchy graph in Fig. 2 to a higher level.

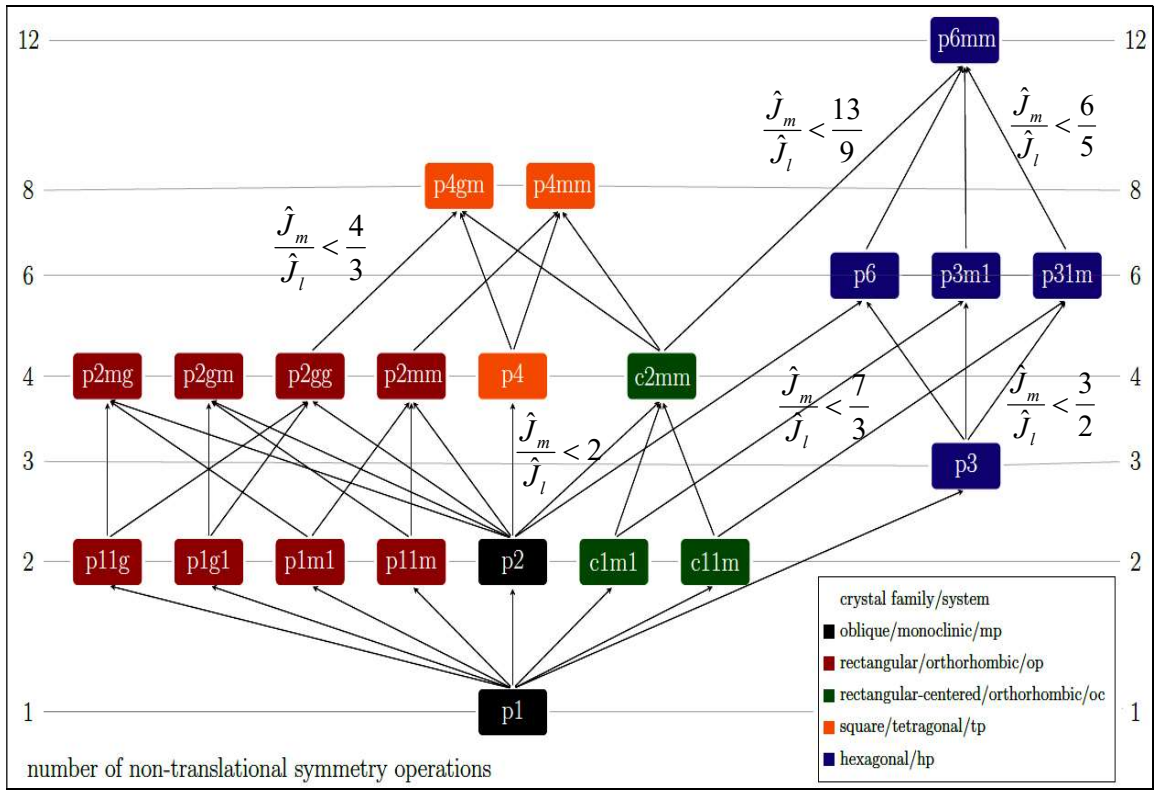

*Figure 2: Applicable hierarchy tree of the plane symmetry groups with limiting ratios of sums of squared complex Fourier coefficient residuals for allowed transitions as insets, from [1]. The hats on top of the J values signify that they need to be estimated from the crystal pattern that is to be classified.*

The *p1* symmetrized geometric model of the crystal pattern that is to be classified is located at the bottom of the hierarchy graph in Fig. 2. It is obtained by Fourier filtering, features strictly enforced translation symmetry, and possesses a squared residual of zero. (The zero-residual value of this model is the reason why there is no inequality for climbing up from the first level of the hierarchy graph in Fig. 2.)

The "anchoring group" [1] features the lowest (non-zero) squared residual $J_l$ (where subscript $l$ stands for *l*ess symmetric) amongst all geometric models. It will with necessity be found at the second or third level of this graph whenever there is more than broken translation symmetry in a real-world crystal pattern. When all models of said pattern at the second and third hierarchy level in Fig. 2 feature comparable and rather high squared residuals, there is probably only broken translation symmetry in the crystal pattern. It is then to be classified as belonging to the *p1* symmetry group.

If the ratio of the calculated $J_m$ to $J_l$ values (where subscript $m$ stand for *m*ore symmetric) for the anchoring group fulfills the inequality that governs the transition from level $l$ to level $m$ in Fig. 2, one can on a preliminary basis conclude that the model that has been symmetrized to the minimal supergroup is in the information-theoretic sense the better representation of the crystal pattern that is to be classified.

For a higher symmetric model (that features a minimal supergroup of the anchoring group) to represent the crystal pattern better (in the information-theoretic sense), the inequalities in Fig. 2 have to be fulfilled for all maximal subgroups (and all of their respective maximal subgroups) simultaneously. If that is not the case, the minimal supergroup designates only a strong pseudosymmetry.

The above-outlined climbing-up testing procedure makes, thus, a very clear/quantitative distinction between genuine symmetries and pseudosymmetries in a crystal pattern. All genuine symmetries must be traceable to the anchoring group by the fulfillment of all of the applicable ratio inequalities in Fig. 2. This is not so for the pseudosymmetries. For them, the corresponding geometric models feature relatively low squared residuals with respect to the Fourier-filtered (*p1*) representation of the crystal pattern that is to be classified, but do not fulfill all applicable ratio inequalities and are, therefore, not traceable to the anchoring group.

## Objectively Classified 2D Translation Periodic Pieces of Graphic Art

The left-hand side of Fig. 3 shows a digital copy of Hans Hinterreiter's beautiful piece of graphic art with the title *4*/ 82$L_{II}$ + 84$E_{II}$ + 42$C_{II}$ + 72$A_2$*. The anchoring plane symmetry group of this crystal pattern is *c11m* because the three vertical mirror lines in the rectangular-centered unit cell [11] are the least broken despite missing white dots, see auxiliary lines on the upper right-hand side of this figure.

The squared residuals of the corresponding non-disjoint pairs of geometric models are such that one is allowed to climb up from *c11m* to *c2mm* but not to *p31m*, see Fig. 2. Upward transitions from both the *p2* and *c1m1* symmetrized models to the *c2mm* model are also allowed by the fulfillment of the applicable ratio inequality in Fig. 2. The ratios of other squared residuals are such that no further climbing up from the *c2mm* symmetrized model, see the upper-right-hand side of Fig. 3, to the *p4gm*, *p4mm*, or *p6mm* symmetrized models of this crystal pattern is permitted. The information-theoretic plane symmetry classification settles for this crystal pattern, therefore, to plane symmetry group *c2mm*.

The *c2mm* symmetrized model features the lowest G-AIC value and the *c11m* symmetrized model possesses the lowest (non-zero) squared residual for all of the geometric models of this crystal pattern. The squared residuals for all of the other geometric models that feature a plane symmetry group on the second and third level of the hierarchy in Fig. 2 are approximately 6 to 8 times larger than that of the model that has been symmetrized to the anchoring group. There are, accordingly, no noteworthy pseudosymmetries in this crystal pattern by Hans Hinterreiter. The techniques that the artist employed to create the graphic artwork on the left-hand side of Fig. 3 were such that a sufficiently similar breaking of the symmetry operations *2*, *.m.*, and *..m* resulted so that all three of them combine to form plane symmetry group *c2mm*.

The lower part of the right-hand side of Fig. 3 illustrates the hexagonal metric of this crystal pattern when a primitive unit cell [11] is used. (While this is permitted, it does not capture the full plane symmetry and should, therefore, be avoided.) Within measurement errors, the magnitudes of the ***a***- and ***b***- axis vectors of the primitive unit cell are equal and the angle between these two vectors is 120°. This feature has been inherited from the "hexagonal grid" that the artists used in the creation of this crystal pattern. The origin of the unit cell is in both of the subfigures fixed to a point with site symmetry *2mm*.

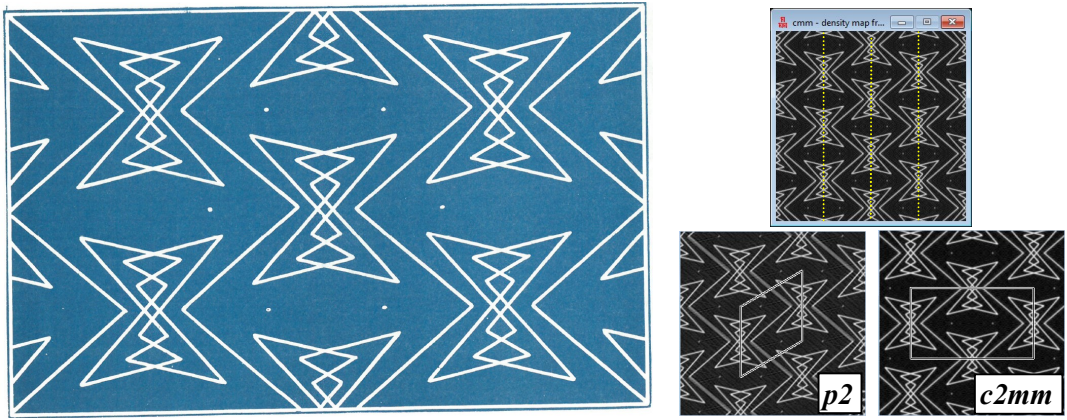

**Figure 3** *Left: Reproduction of Hans Hinterreiters's 4\*/ 82L$_{II}$ + 84E$_{II}$ + 42C$_{II}$ + 72A$_2$ as shown in [12], but here in color), comprising approximately three rectangular-centered (or six primitive) unit cells. **Upper right:** This pattern's c2mm symmetrized gray-level model as obtained after Fourier-back transforming to direct space with vertical mirror lines marked by dotted auxiliary lines. **Lower right:** A primitive and a rectangular-centered gray-level model of this crystal pattern with one unit cell outlined for each of the two non-disjoint geometric models of the crystal pattern to their left.*

The left-hand side of Fig. 4 shows a digital copy of a section of Eva Knoll's charming collage with the title *Tiles with quasi-ellipses*. The upper right-hand side of this figure shows a copy of the photograph of the painted tile that was used to create this collage by artful tessellation. The lower right-hand side of Fig. 4 shows a hand sketch of an earlier design idea for this kind of crystal pattern. (Note in passing that the painted tile does not follow this sketch exactly with respect to its central brown square.)

Human classifiers are bound to assign plane symmetry group *p4gm* to the crystal pattern on the left-hand side of Fig. 4, at least at first sight. (The author did so as well originally.) This is because four-fold rotation points, two-fold rotation points, and glide as well as mirror lines are all visually recognizable in mutual orientations that facilitate this classification. In a *qualitative* sense, all of these symmetries combine to plane symmetry group *p4gm*. The human tendency to overestimate plane symmetries in noisy crystal patterns when strong pseudosymmetries are present is presumably a consequence of how symmetry hierarchies are perceived by means of the human visual system [13]. The presence of strong (but unrecognized) pseudosymmetries tends to create "confusion" about what is to be concluded from the applicable hierarchy of the plane symmetry groups whenever classifications are made by sight only or, in other words, subjectively.

The objective classification with the author's method reveals, on the other hand, *p2* as the anchoring group with an allowed transition to plane symmetry group *p4* only. The geometric models of this crystal pattern that were symmetrized to plane symmetry groups *p11g*, *p1g1*, *c1m1*, and *c11m* all possess very low squared residuals that individually allow for transitions to the *p2gg* and *c2mm* symmetrized models at the fourth hierarchy level in Fig. 2. Climbing up from the geometric model that features the anchoring group to the *p2gg* and *c2mm* symmetrized models is, however, forbidden because the corresponding ratios of squared residuals are too large to fulfill the inequality limit for transitions from the second to the fourth level. The climbing up from the *p2gg* and *c2mm* symmetrized models to the *p4gm* model is also allowed, but the latter is also only a strong pseudosymmetry as it cannot be traced to the anchoring group by all of the applicable three routes.

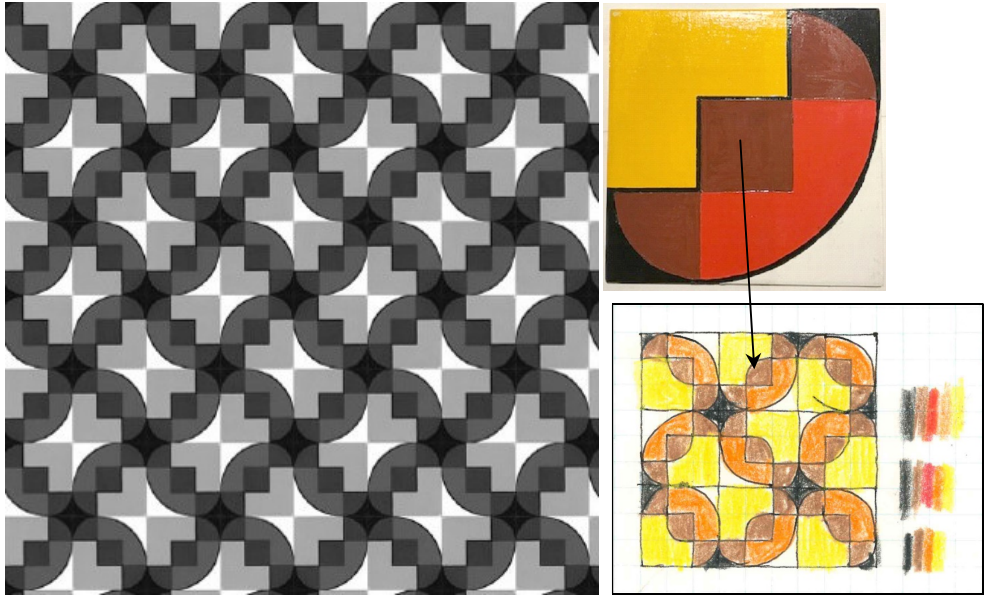

**Figure 4** *Left: Gray-level reproduction of Eva Knoll's Tiles with quasi-ellipses (1992, acrylic on ceramic), modified from [1], showcasing plane symmetry group p4 (rather than p4gm) and several strong pseudosymmetries. **Upper right:** Copy of the color photo of the hand-painted tile (on a white sheet of paper). **Lower right:** Hand sketch of a related design by the artist. The arrow from the upper to the lower part of this subfigure has been added in order to illustrate that one copy of the hand-painted tile represents one-quarter of a translation periodic unit cell of the crystal pattern.*

This rather surprising result can be fully understood from the sequence of creative processes that went into the creation of Eva Knoll's crystal pattern on the left-hand side of Fig. 4. The artist painted a single asymmetric unit [14] onto a single ceramic tile by hand, upper right-hand side of Fig. 4. The painted asymmetric unit features a slightly broken mirror line across one of its two diagonals and covers the whole ceramic tile. That tile has a square shape to a very good approximation and is 6 inches long on its edges. The artist took a color photo of that painted tile and produced multiple copies of a much smaller photo with the shape of a square of the same size.

Sets of four photos of the tile were assembled into four-fold larger squares with four-fold rotation points at their centers by making sure that the slightly broken mirror lines run along the fractional coordinates $x, x+½$, $-x, -x+½$, $-x+½, x$, and $x+½, -x$ of the thus created square unit cell. It is quite remarkable that three pairs of slightly broken glide lines were created in the unit cell as a result of this highly creative assembly process. As the lower right-hand side of Fig. 4 demonstrates, the specifics of the assembly were according to an earlier creative design plan.

The so-created (four-fold larger) unit cell squares were then laid out on a square 2D lattice without overlaps or gaps. This created four-fold rotation points at each of the four vertices of the unit cell and two-fold rotation points in the middle of each of its four edges. The whole piece of Eva Knoll's graphic artwork consists, thus, of a translation periodic array of four properly assembled photocopies of her painted tile (asymmetric unit). The graphic artwork features plane symmetry group *p4* as the result of its creation process.

The genuine site symmetries in the assembly are point groups *4* and *2*, which are non-disjoint. The artistically sophisticated distribution of paint, the slightly broken mirror line in the original asymmetric unit, and the two- and four-fold rotation points that resulted from the translation-periodic assembly process combined to the above-mentioned strong pseudosymmetries. It is obviously nearly impossible to create a mirror line by hand in a painting that is broken to such a small amount that its symmetry breaking becomes comparable to the precision of the industrial production process of the ceramic tile that was hand painted.

## Summary and Conclusions

This paper explained the author's objective plane symmetry group classification method very briefly and applied it to two beautiful pieces of graphic art. Genuine symmetries that combine to the derived plane symmetry group with site-multiplicity four were distinguished from strong pseudosymmetries for one of these artworks. Human classifiers would most likely overestimate the plane symmetry group in this particular piece of graphic art by failing to make such a distinction. This fact can be taken as a testament to the accuracy, precision, and effectiveness of the new method. As structural crystallography is loosely speaking about averaging over correctly identified asymmetric units in experimental data, this method is bound to impact the future practice of how that science is conducted. It might, perhaps, eventually also be taken up by art critics, librarians, and graphic artists when objective plane symmetry classifications and quantizations are desired. The other analyzed piece of graphic art inherited a hexagonal metric at the primitive unit cell level from the particulars of its creation process, but did not feature recognizable pseudosymmetries. Its symmetry classification by a human being based on its visual appearance is just the same as that obtained by the information-theory based method.

## Acknowledgments

The author is grateful for proofreads by the current student members of his research group: Tyler Bortel and Benjamin Fischer-Shane. Professor emeritus Emil Makovicky of the Department of Geosciences and Natural Resource Management of the University of Copenhagen is thanked for a digital version of the analyzed artwork by Hans Hinterreiter, left-hand side of Fig. 3. Professor Eva Knoll of the Department of Mathematics of the University of Quebec at Montreal is thanked for both discussions around her crystal pattern and her contributing an appendix to the expanded *arXiv* version of [1] on how she created that pattern. That appendix served as the source for the two subfigures on the right-hand side of Fig. 4.